# The Complex Brain Hypothesis:
# Resolving the Entropy-Content Conundrum in Minimal Phenomenal Experience

Jonas Mago[1+*], Edmundo Lopez-Sola[2,3+], Jakub Vohryzek[2,3+], Michael Lifshitz[4,5], Robin Carhart-Harris[6,7], Karl Friston[8], Shamil Chandaria[2]
[+] Shared first authorship
[*] Corresponding author: jonas.h.mago@gmail.com

[1] Integrated Program in Neuroscience, McGill University, Montréal, Quebec, Canada
[2] Centre for Eudaimonia and Human Flourishing, University of Oxford, UK
[3] Center for Brain and Cognition, Universitat Pompeu Fabra, Barcelona, Spain
[4] Division of Social and Transcultural Psychiatry, McGill University, Montréal, Québec, Canada
[5] Lady Davis Institute for Medical Research, Jewish General Hospital, Montréal, Québec, Canada
[6] Departments of Neurology, Psychiatry and Behavioral Sciences, Weill Institute for Neuroscience, University of California San Francisco, San Francisco, CA, USA
[7] Centre for Psychedelic Research, Imperial College London, London, UK
[8] Department of Imaging Neuroscience, Queen Square Institute of Neurology, University College London, 12 Queen Square, London WC1N 3AR, UK

**Abstract**

Minimal Phenomenal Experiences (MPEs) are states of consciousness in which wakefulness is preserved but phenomenal content is low or absent. The Entropic Brain Hypothesis (EBH) is a model of conscious processes that regards the entropy of spontaneous brain activity as a marker of 'phenomenal richness', exemplified by high-content psychedelic experiences (HCPEs). Yet recent human neuroimaging studies of MPEs induced by meditation—and possibly 5-MeO-DMT—suggest that these states, defined by their phenomenological simplicity, also show signs of increased neurophysiological entropy. This presents a conundrum for the EBH: brain entropy is elevated with increased and decreased richness of the phenomenal experience. Here, we put forward the Complex Brain Hypothesis (CBH), which proposes that the richness of experience differentiating MPEs from HCPEs is better indexed by complexity than by entropy. We argue that brain complexity is modulated by the grain of inference through which the brain resolves uncertainty: some HCPEs exemplify a fine-grained regime, in which loosened constraints amplify fluctuations into proliferating content, whereas some MPEs exemplify a coarse-grained regime, in which a simpler model dissolves variety into an experience of 'contentless' awareness. Both regimes can be associated with elevated brain entropy, but they diverge in phenomenology and perturbational signatures. By resolving the entropy–content conundrum, the CBH refines the EBH and highlights MPEs as an important test case for computational theories of consciousness.

## Introduction

The relationship between neural complexity and conscious experience is a central challenge in neuroscience (Seth & Bayne, 2022). A common assumption, crystallized in the "entropic brain hypothesis" (EBH), is that the entropy of spontaneous brain activity scales monotonically with the phenomenological richness of experience, within lower and upper limits, beyond which brain entropy would not be associated with consciousness (Carhart-Harris, 2018, 2025; Carhart-Harris et al., 2014; Carhart-Harris & Friston, 2019). In this view, richer, more elaborate experiences are accompanied by more entropic (i.e., high average information) neural dynamics, while states of reduced awareness (e.g., deep sleep, anesthesia) show a more ordered (low-entropy) brain state with predictable patterns of brain activity and a matching low-content or absent quality of consciousness (Bocaccio et al., 2019; DiNuzzo & Nedergaard, 2017; Gervais et al., 2023; Meisel et al., 2017; Toker et al., 2022, 2024). The core intuition that grounds the EBH can be understood in neurophenomenological terms (Ramstead, Seth, et al., 2022; Sandved-Smith et al., 2025; Varela, 1996) as the following argument and reasoning:

1. First person phenomenology, conscious experience, has informational content; e.g., we see a coin facing heads up not tails – this is at least 1 bit of information contained in experiential content.

2. Richer and more elaborate phenomenal experiences therefore have more informational content.

3. Information processing in the brain is associated with the informational contents of experience.

4. Richer and more elaborate phenomenal experiences will be associated with increased potential information (i.e., higher entropy) carried by neuronal dynamics.

The last assumption entails a strong commitment to the notion that the entropy of (usually neurophysiological) brain states reflect the entropy of representations or experiences encoded by those states. The implicit distinction foregrounds a key issue when talking about brain entropy and complexity. There is a fundamental difference between the entropy of brain states and the entropy of probabilistic beliefs entailed by brain states. Under the free energy principle, this can be read as a distinction between the (intrinsic) information geometry afforded by the thermodynamics of neuronal activity and the (extrinsic) information geometry associated with the probabilistic beliefs encoded by neuronal activity [2]. On this view, there is a close relationship between the entropy of brain activity and the entropy of beliefs encoded by brain activity that arises in several settings. For example, the principles of efficient encoding suggest that the variational and thermodynamic free energy share the same minima (Sengupta et al., 2013, 2016). Free energy corresponds to entropy under some constraints supplied by an energy or potential (Jaynes, 1957; Klein & Meijer, 1954; Seifert, 2005; Tomé, 2006). This means that minimizing free energy corresponds to maximizing entropy, under constraints (Jaynes, 1957, 1989; Sakthivadivel, 2022, 2023). Variational free energy pertains to the entropy of beliefs, while thermodynamic free energy refers to the entropy of brain states encoding those beliefs. The close relationship between variational and thermodynamic free energy (and entropy) is reflected in the physics of information; such as Landauer's principle (Bennett, 2003; Landauer, 1961) and the Jarzynski equality (Evans, 2003; Jarzynski, 1997). In terms of consciousness studies, the close relationship leads to a dual aspect (Markovian) monism, in which the variational and thermodynamic entropy rest upon the same dynamics [2].

Crucially, one can rearrange the constraints (i.e., expected energy) and entropy in free energy to express it as complexity minus accuracy (Penny, 2012). Complexity on this view scores the divergence between posterior and prior beliefs (see figure 1). In other words, it reflects the degrees of freedom used to provide an accurate account of the sensorium and thereby reflects the requisite law of variety (Ashby, 1956). We will appeal to this notion of complexity later. Interestingly, in the context of neuropsychology, complexity has been associated with redundancy, while entropy corresponds to degeneracy (Sajid et al., 2020; Tononi et al., 1999). The free energy principle and entropic brain hypothesis can therefore be read as celebrating degenerate representations in the sense of Laplace's principle, Occam's principle and the maximum entropy principles that attend the physics of measurement (Jaynes, 1957; Maisto et al., 2015).

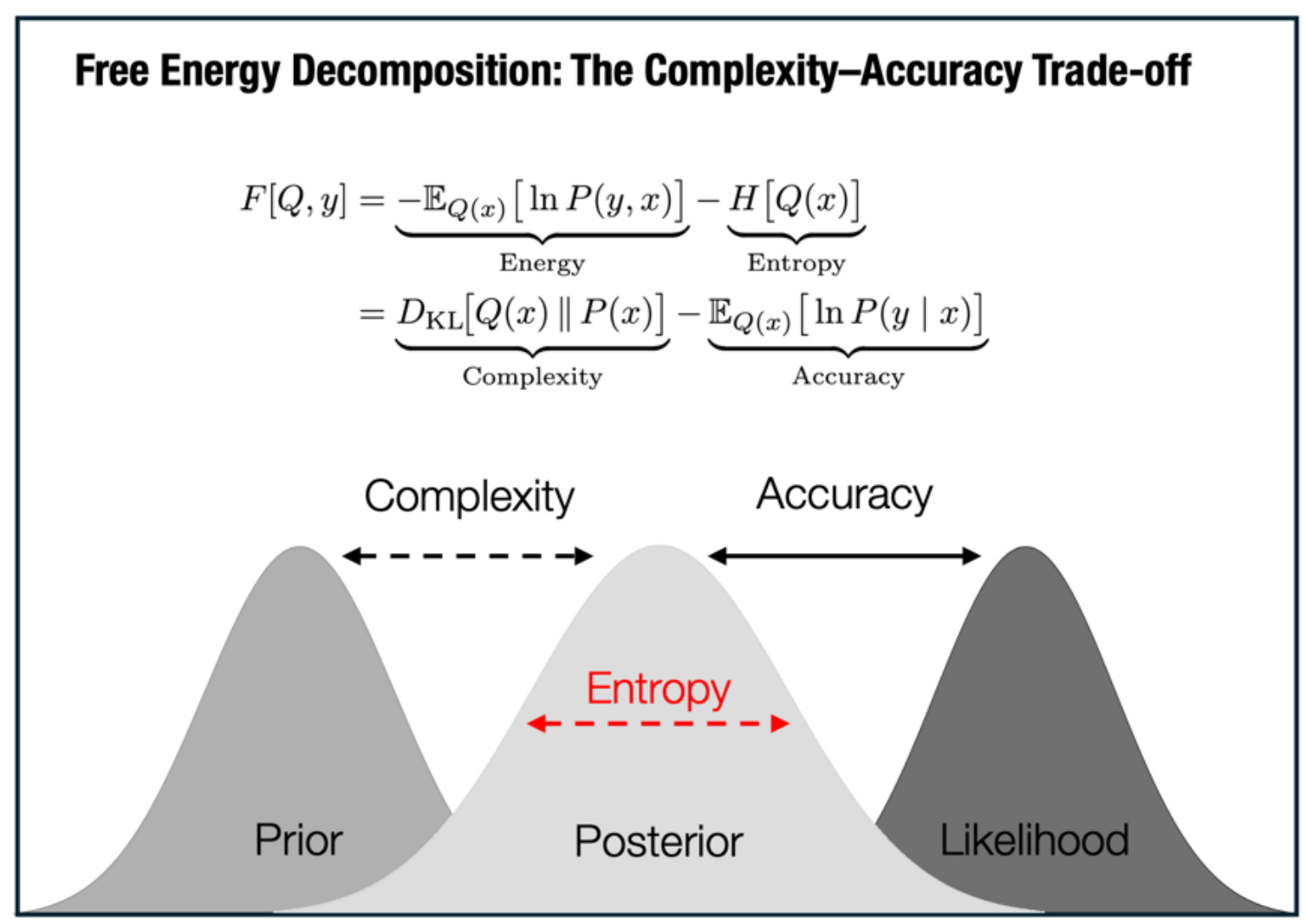

**Figure 1. Free energy as a trade-off between complexity and accuracy.** The variational free energy F can be decomposed into a complexity term and an accuracy term. Complexity corresponds to the Kullback–Leibler divergence between posterior and prior beliefs, $D_{\{KL\}}[Q(x)||P(x)]$, and reflects the degree to which posterior beliefs must deviate from prior expectations to account for sensory data. Accuracy corresponds to the expected log-likelihood, $\mathbb{E}_{Q(x)}[ln\, P(y|x)]$, and reflects how well posterior beliefs explain the observations. In the formal expression, free energy is given by complexity minus accuracy. Because log-likelihoods are typically negative, the subtraction of accuracy is equivalent to adding prediction error or inaccuracy. Minimizing free energy therefore entails a trade-off: posterior beliefs should remain close to the prior (low complexity) while still providing an accurate account of sensory input (high accuracy). The posterior thus occupies a compromise position between prior expectations and the likelihood imposed by sensory evidence.

A complementary view of free energy minimization can be read as complexity minimization, under (accuracy) constraints (Hinton & Van Camp, 1993; Hinton & Zemel, 1993). Complexity minimization emerges in the context of algorithmic complexity through minimum description or length formulations (Ruffini, 2017; Wallace & Dowe, 1999); in which minimizing complexity corresponds to maximizing compression (Ruffini, 2017; Schmidhuber, 2010). This formulation underwrites treatments of universal computation and Solomonoff induction (Hutter, 2005; Solomonoff, 2009) that itself inherits from Kolmogorov complexity (Wallace & Dowe, 1999). In what follows, we now consider the distinction between entropy and complexity in light of various brain states and their measurement.

Taken together, the considerations above motivate the following qualified inference: Shannon entropy is a measure of information content of a probabilistic process or phenomenon, and so, following from premises 1-4 above,  we would expect richer and more elaborate phenomenal experiences to be associated with higher brain entropy measures (e.g., neural signal diversity measures such as Lempel-Ziv complexity, LZc). However, as we will argue here, the converse is not necessarily true, i.e., higher brain entropy does not necessarily imply richer experience, such as in the case of MPEs.

The above could suggest that any high entropy state measured in the brain should be associated with richer and more elaborate experiences. However, Minimal Phenomenal Experiences (MPEs)—states with little or no phenomenal content—provide contradictory evidence. MPEs have recently been proposed as a distinct class of conscious states, characterized by extreme phenomenological simplicity and the reduction, or even absence, of intentional content (Metzinger, 2020, 2024). Unlike ordinary states of awareness, which are structured around objects, thoughts, and sensory inputs, MPEs are marked by a mode of psychological absorption and sensory fading that can culminate in contentless—yet states of high (luminous) awareness (Metzinger, 2024). MPEs present as a self-sustaining mode of bare wakefulness, sometimes called a pure consciousness experience (PCE), rather than a representation of specific objects or events (Forman, 1999; James, 1937). Phenomenological reports suggest that MPEs can be intentionally cultivated through contemplative practice and pharmacological induction (Mago et al., 2025; Sparby & Sacchet, 2024; Timmermann et al., 2025), or can occur spontaneously. Phenomenological surveys indicate that such minimal states are not confined to a single lineage but recur across diverse contemplative traditions (Metzinger, 2024). Crucially, MPEs challenge the assumption that conscious experience is defined by "content", demonstrating that pure awareness can persist even when representational contents are radically attenuated. This makes them a uniquely powerful test case for computational accounts of consciousness.

MPEs have been associated with high neural signal diversity in meditators (Lieberman et al., 2025; Mago et al., 2025; Potash, van Mil, et al., 2025; Potash, Yang, et al., 2025; Shinozuka et al., 2025; Vohryzek et al., 2025). Tentative results also indicate an increase in neural signal diversity and complexity in the trajectory of slow-wave patterns in the context of 5-MeO-DMT (Blackburne et al., 2025; Timmermann et al., 2025), though further work is needed to confirm these findings. These states, whether cultivated through advanced contemplative practice (e.g., absorptive concentration such as jhāna, or cessations) or induced pharmacologically (e.g., 5-MeO-DMT), are typically described in terms of deep absorption, sensory fading, and in some cases near-contentless-ness (Laukkonen et al., 2023; Timmermann et al., 2025; Van Lutterveld et al., 2024).

Thus, we have experiences with minimal content (e.g., jhāna, 5-MeO-DMT) and of phenomenological richness (e.g., LSD, psilocybin, N,N-DMT) that both exhibit high entropy. We refer to this dissociation as the entropy–content conundrum: when measures of brain entropy do not track with the richness of phenomenal content.

To address this emerging gap, we propose the Complex Brain Hypothesis (CBH). According to the CBH, the richness of conscious experience is better indexed by the complexity of the beliefs entailed by a generative model—the brain's predictive 'world model' used to explain and anticipate sensory data—rather than by entropy alone. Technically, the complexity of (Bayesian) beliefs encoded by brain activity is equal to the entropy of those beliefs, relative to some prior beliefs. In other words, complexity is a relative entropy (a.k.a. KL divergence) between posterior and prior beliefs.

In this sense, entropy reflects the total information in—or encoded by—neuronal dynamics, whereas complexity tracks how that information is organized into structured, differentiated contents in relation to prior beliefs in the absence of sensory evidence. As a relative entropy, it scores the degrees of freedom used to explain the sensorium or, intuitively, the degree to which sensory evidence has 'changed (the contents of) one's mind'.

In the sections that follow, we show how distinguishing entropy from complexity resolves the entropy–content conundrum and situate both MPEs and high-content psychedelic experiences (HCPEs) within a unified framework.

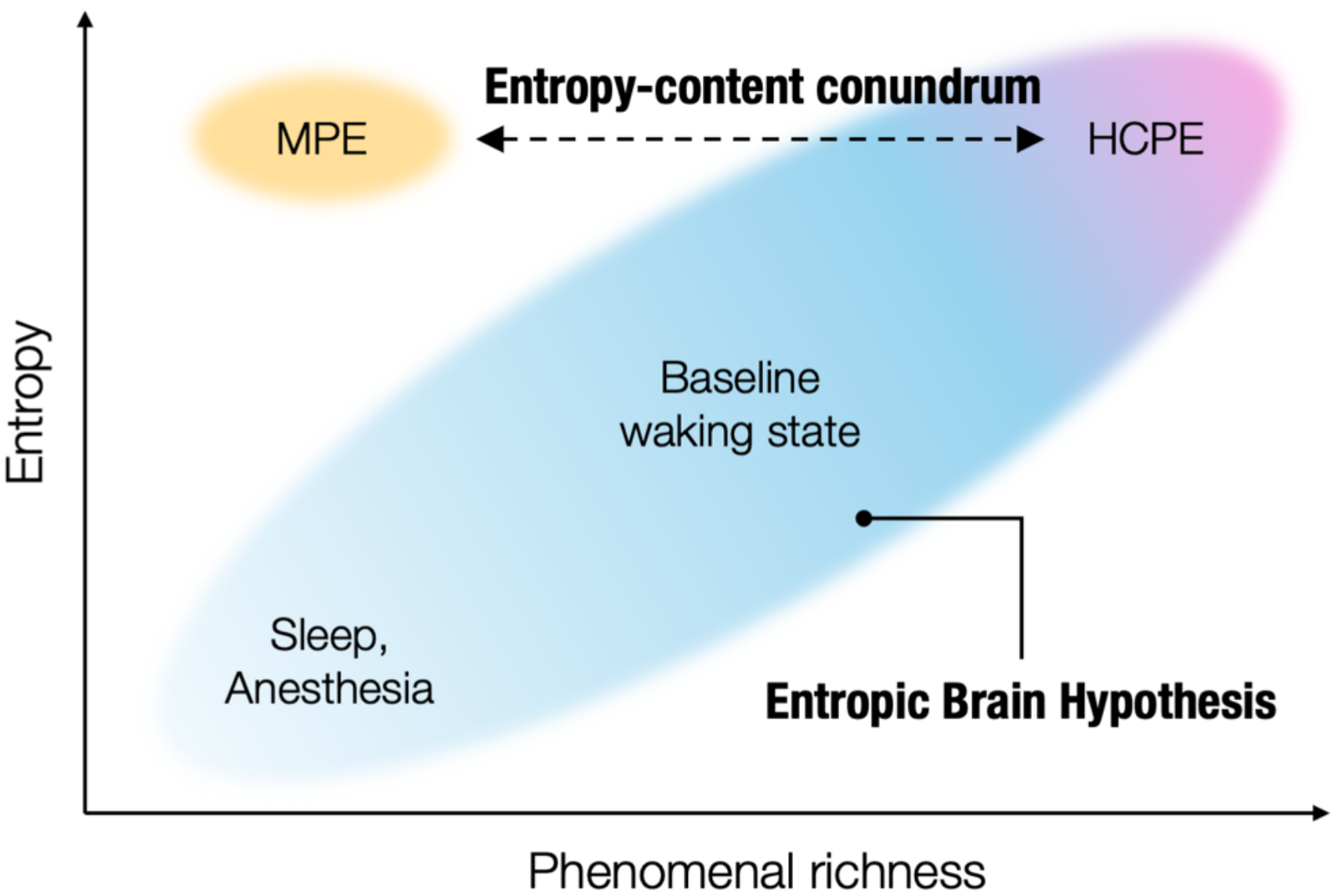

**Figure 2. The entropy–content conundrum.** The entropic brain hypothesis (EBH) implies a monotonic mapping from phenomenological content to entropy. This would predict that minimal-content states are associated with low entropy while content-rich states are associated with high entropy. However, empirical evidence indicates that both MPEs (e.g., jhāna, 5-MeO-DMT) and HCPEs (e.g., LSD, psilocybin, N,N-DMT) exhibit elevated entropy relative to ordinary wakefulness, challenging a simple monotonic relationship between phenomenological richness and brain entropy.

**Two Inferential Regimes**

We propose that elevated entropy can arise under two distinct inferential regimes. Some psychedelics (LSD, psilocybin, N,N-DMT) are theorized to relax the precision of high-level priors, increasing the influence of sensory evidence on perceptual inference (Brouwer & Carhart-Harris, 2021; Carhart-Harris et al., 2014, 2016; Carhart-Harris & Friston, 2019). The loss of precise prior constraints (i.e., inducted biases) renders the generative model overly accommodating, readily explaining all sensory fluctuations as a meaningful signal. This yields a generative model with unconstrained degrees of freedom, which is prone to overfitting sensory data, dynamically metastable, very sensitive to perturbations, and characterized by rapid configuration changes. The phenomenology is one of overflow: vivid, multifaceted (elemental) content unleashed by loosened constraints.

In terms of Bayesian belief updating, a loss of prior precision corresponds to an increase in the entropy of (high level) beliefs, while unattenuated sensory prediction errors drive posterior beliefs away from prior beliefs, engendering rich perceptual content and high complexity. In terms of perceptual inference, this would correspond to overfitting the sensorium.

By contrast, absorptive meditative states and 5-MeO-DMT can reveal an experience that is better characterized by phenomenological 'simplicity' or 'stillness'. For example, in jhāna, the meditative object saturates awareness, such that peripheral inputs are attenuated and exert little influence; in 5-MeO-DMT, the system can lock into a uniform experiential mode resistant to perturbation. In both cases, there is an increase in brain entropy, but ascending sensory prediction errors are not attended to (in the case of jhāna through maintaining a certain attentional set). Sensory prediction errors persist, but they are not expressed as proliferating content, because their precision is attenuated.

In terms of Bayesian belief updating, a loss of prior precision corresponds to an increase in the entropy of (high level) beliefs, while attenuated sensory prediction errors are unable to drive posterior beliefs away from prior beliefs, engendering low complexity. In terms of perceptual inference, this would correspond to under fitting the sensorium. See Figure 3 for a schematic illustration of this dissociation.

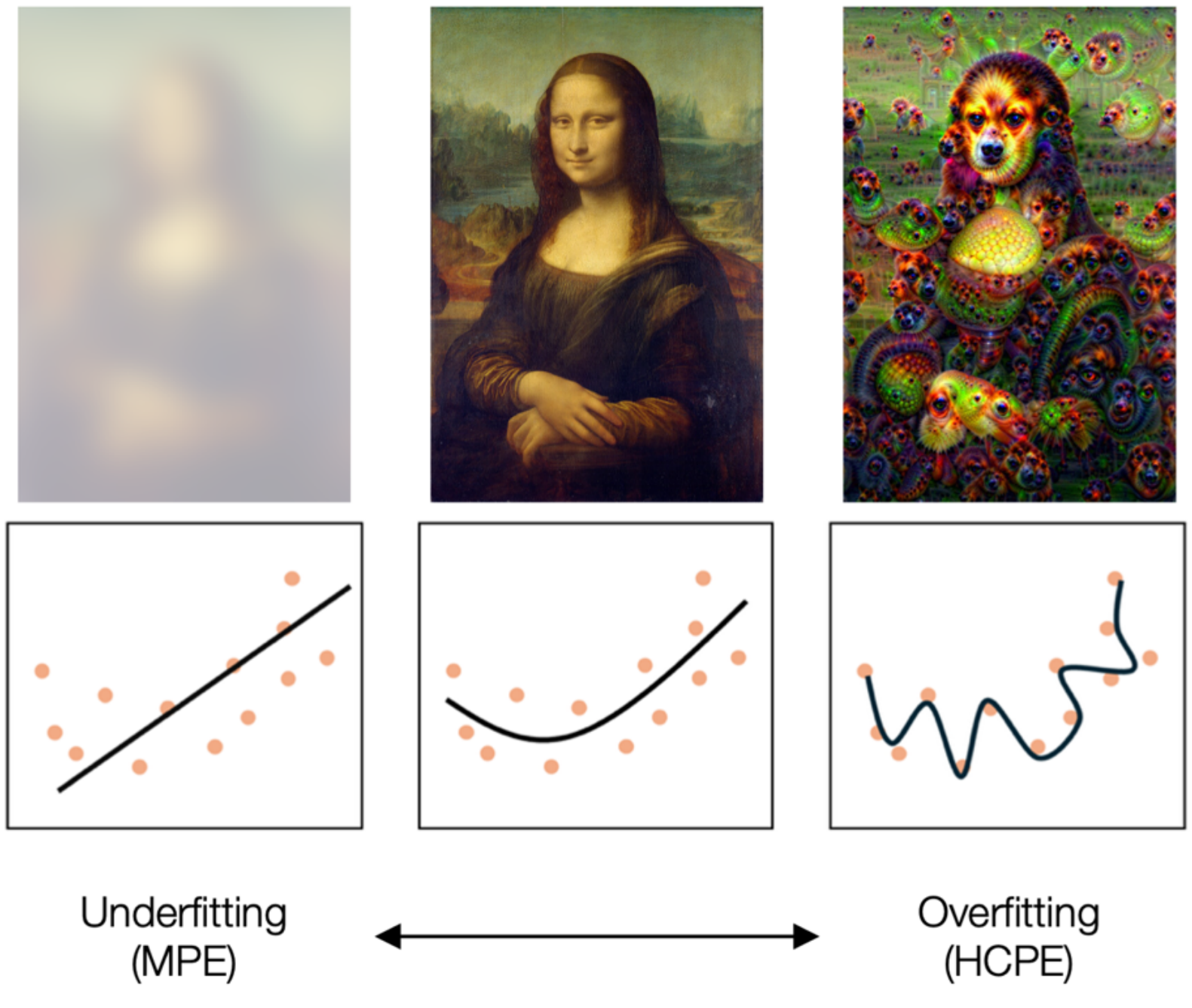

**Figure 3. Two inferential regimes.** A conceptual under/overfitting illustration. HCSPs map to overfitted inference, while absorptive states such as jhāna and 5-MeO-DMT map to underfitted inference. Left: an underfitting regime (our MPE case) smooths over variation, producing a vague, content-minimal percept. Center: an intermediate, well-regularized model yields the familiar percept of Leonardo da Vinci's Mona Lisa (public domain image). Right: an overfitting regime (HCPEs) treats small fluctuations as meaningful (precise) structure and elaborates them into a densely patterned image. The image was generated with Google's DeepDream algorithm (Mordvintsev et al., 2015) and is sourced from Wikimedia Commons ("Mona Lisa" with DeepDream effect using VGG16 network trained on ImageNet, 2024).

These two inferential regimes can also be understood through the lens of algorithmic information theory (AIT). A central insight of AIT is that the best explanation of a dataset is the one that provides the greatest compression. This is captured by the notion of Kolmogorov complexity, which is the length of the shortest program that can reproduce the dataset in a Universal Turing Machine (Hinton & Zemel, 1993; Liu et al., 2008; Wallace & Dowe, 1999):

$$K(v) = \min_{p} L(p): U(p) = v$$

where v is the sensory data, L(p) is the length of the program which can be denoted by a binary number p, and U(p) is the output of a Universal Turing Machine running the program. In words the formula says roughly that the Kolmogorov complexity is the minimum length of some program p such that when that program is run on a universal Turing machine U it will output v. We can think of the brain as attempting to discover such a minimal program that most efficiently compresses its sensory data (Ruffini, 2017; Ruffini et al., 2024).

While Kolmogorov complexity defines a theoretical ideal of simplicity, the Minimum Description Length (MDL) principle operationalizes this insight in statistical terms: the optimal model is the one that minimizes the total description length of some (sensory) data D

$$L(D) = L(model) + L(D|model) = -\log P(D)$$

that is, the information required to specify both the model, L(model), (its effective number of parameters or model complexity) and the residual error or unexplained variability, L(D | model) (Liu et al., 2008). Interestingly, Bayesian inference can be viewed as a process of minimizing description length (Grünwald, 2007), balancing model complexity and explanatory accuracy.

As noted above, this formulation naturally connects with active inference under the free-energy principle (Friston, 2009, 2010; Friston et al., 2006) and variational inference generally (Hinton & Zemel, 1993). Technically, variational free energy is a tractable bound on total description length: where the negative description length L(D) is known as log model evidence or marginal likelihood in statistics, or an evidence lower bound (ELBO) in machine learning (Winn et al., 2005). It is important to note that the description length is not computable (as with most constructs in AIT). This is why variational free energy is used as a (computable) bound on the marginal likelihood L(D) in statistics, machine learning and—under the free energy principle—the brain.

Here, L(model)—i.e., complexity—refers to the description length of the brain's generative world model or—in the current setting—the generative phenomenal unified world model or "reality model" (Laukkonen et al., 2025). In active-inference terms, hierarchical inferences are Bayesian-bound into a global posterior that is homologous to this reality model; this global posterior sets the space of what can be known or becomes conscious (i.e., the phenomenal field). Thus, the "model" in L(model) is the organism's posterior generative model of the world, the very model whose contents partially are the contents of experience. Moreover, because this reality model is reflexively broadcast through the hierarchy, aspects of the posterior are available as awareness, providing a principled bridge from computational description length (MDL) to phenomenology.

In this formulation, $L(D) = -\log P(D) < F(D)$ corresponds to the shortest possible description of the data, where variational free energy $F(D)$ provides an upper bound on this quantity. Because L(D) is also the self-information of the data under a generative model, its average corresponds to the entropy of the data. Therefore, minimizing variational free energy minimizes the entropy of data. Note, the entropy of the data is not the entropy of Bayesian beliefs about the hidden causes of data or the neuronal parameterization of those beliefs, which is maximized (Da Costa et al., 2021; Jaynes, 1989; Ramstead, Sakthivadivel, et al., 2022).

From the perspective of Bayesian mechanics (and AIT), the entropy–content conundrum can be resolved in the following way. Under HCPEs, the brain effectively increases the degrees of freedom available to explain sensory input, i.e., model complexity L(model) increases, expanding the repertoire of posterior beliefs it uses to explain or infer bottom-up data. As outlined in the reasoning underlying the EBH, the richness of phenomenological content can be associated with the complexity of the perceptual inference. Accordingly, this regime manifests phenomenologically as a proliferation of content and sensory detail. The high entropy due to a relaxation of prior precision is thereby accompanied by increasing model complexity.

Conversely, in MPEs (as in jhāna or under 5-MeO-DMT), both prior and sensory precision are attenuated; thereby reducing the degrees of freedom used in explaining the sensorium, resulting in a high-entropy, low-complexity perceptual inference (where the posterior does not diverge from the prior due to the absence of sensory entailment). Phenomenologically, the low complexity corresponds to a particular experience with minimal content. This underfitting regime captures little of the sensory variability, because it is free from the accuracy constraints of L(data|model); i.e., data are deemed imprecise lose their ability to update prior beliefs. The variational free energy or total description length, L(data), is still minimized, but complexity minimization dominates.

Thus, entropy remains high in both inferential regimes, but the phenomenological expression differs: HCPEs amplify perceptual complexity, while MPEs suppress it, yielding minimal-content awareness.

The minimization of complexity or compression can also be intuited in terms of coarse-graining. Coarse graining can be implemented not only by reducing the number of representational degrees of freedom, but also by changing their precision. In predictive coding terms, precision determines how strongly prediction errors update posterior beliefs across the hierarchy. Lowering prior precision effectively collapses the space of distinguishable hypotheses: many micro-variations in the signal are treated as irrelevant noise, and only a small set of coarse, global features remains behaviorally and phenomenologically “expressed.” In this sense, precision acts as a soft form of coarse-graining: even if the underlying model architecture is unchanged, changes in precision can prune the model’s effective parameterization (and thus its apparent complexity), shifting inference toward an underfitting regime in which experiential content is smoothed away while unresolved variability can still keep entropy high.

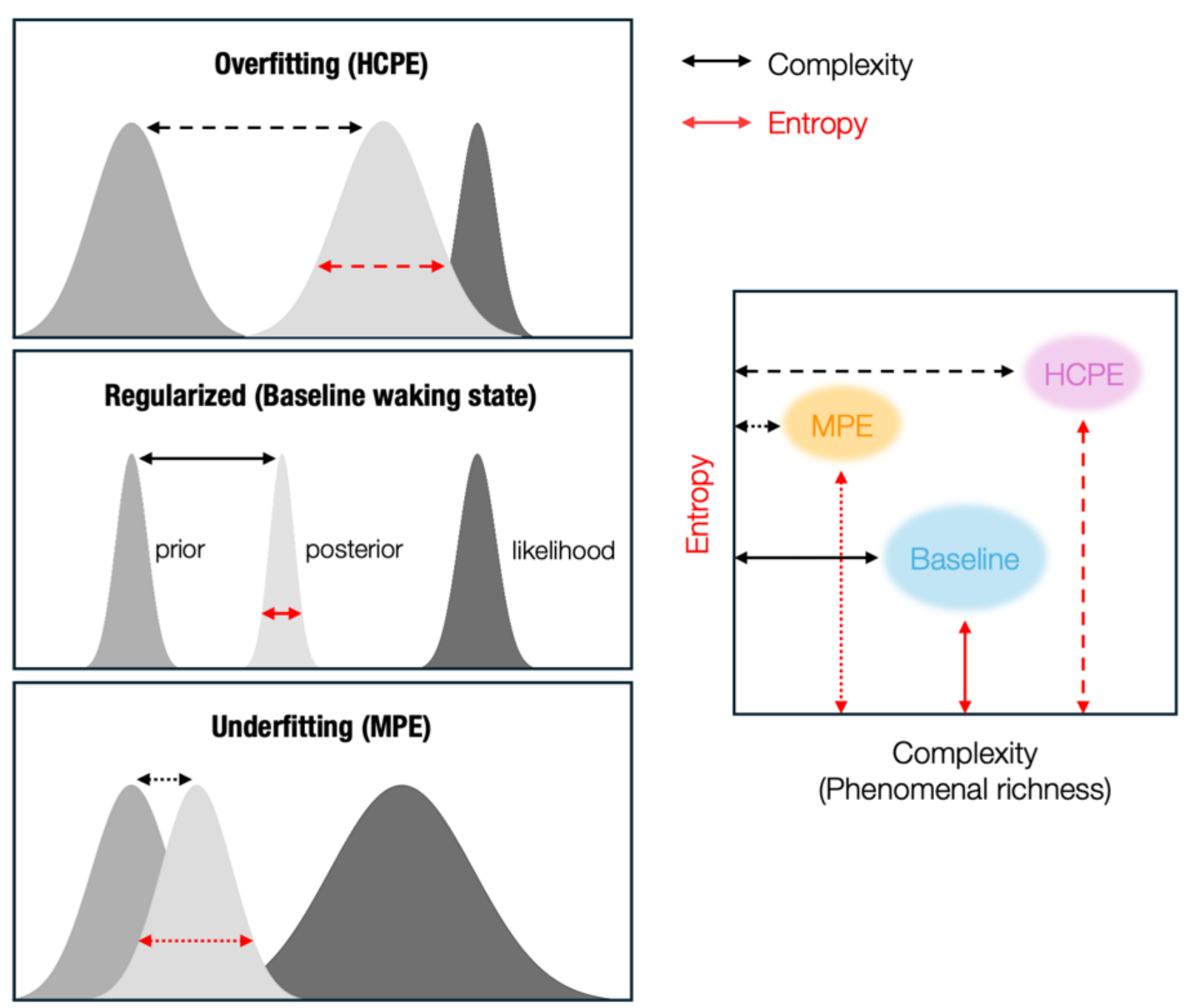

**Figure 4. Solving the conundrum: complexity, entropy and phenomenal richness.** Conceptual illustration of the relationship between model complexity (black), entropy (red), and phenomenal richness. The left panels recapitulate the three regimes of inference in Figure 3 that inherit from reduced prior precision in the context of unattenuated and attenuated sensory (i.e., likelihood) precision (top: overfitting and bottom: underfitting, respectively). The complexity corresponds to the divergence between the prior and the posterior, while the entropy corresponds to the dispersion of the posterior. These schematic depictions of complexity and entropy enable us to revisit Figure 2 (right panel) and resolve the entropy-content conundrum by associating the phenomenal richness of content with complexity.

## Apparent Complexity and Coarse-Graining

The relationship between phenomenal content and complexity is fundamentally shaped by the nature of the generative model that, effectively, carves nature at its joints. This can be considered as a progressive abstraction or coarse-graining as one ascends levels in—or scales —in a hierarchical generative model [this is sometimes cast in terms of the renormalization group: (Friston et al., 2025; Hu et al., 2022; Mehta & Schwab, 2014; Watson et al., 2022)]. In brief, the same physical system can look highly structured at one level of description yet trivially simple at another. To illustrate this scale-dependence, we introduce a simple example.

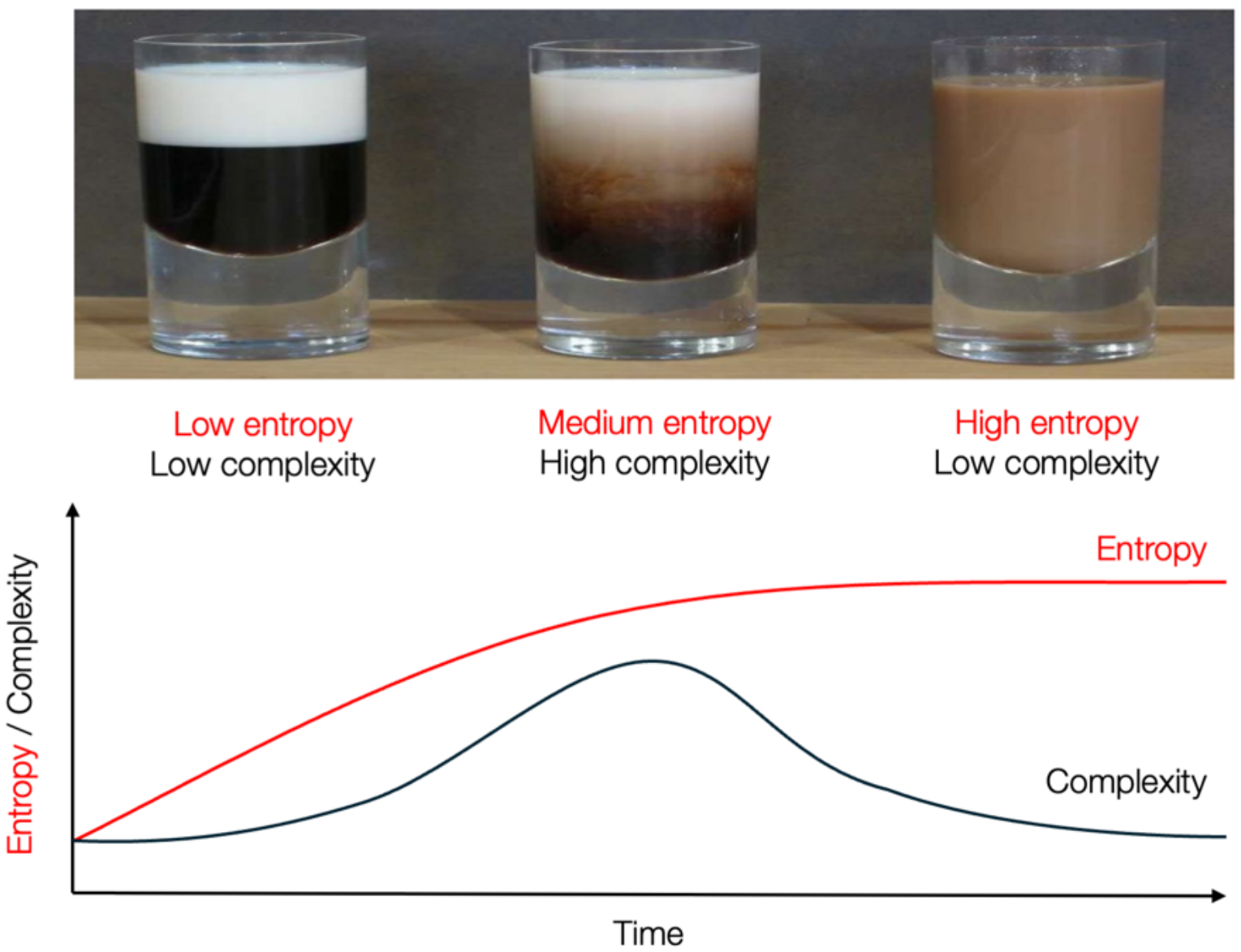

**Figure 5. Entropy and complexity in the coffee–cream mixing process (adapted from Aaronson et al., 2014)**. As black coffee and white cream mix, entropy increases monotonically, reflecting the growing number of particle configurations consistent with the systems state. At coarse-grained scales, however, complexity rises during the formation of structured swirls and then collapses once the mixture becomes uniform, illustrating that high entropy does not imply high complexity. This scale-dependence provides a simple intuition for how the same system can appear richly structured or phenomenologically simple depending on the granularity of the model used to describe it.

Consider a cup of coffee which starts as black coffee with white cream floating on top (left cup). If this is modelled as black and white particles, as time proceeds the particles mix and the entropy of the system increases monotonically, initially creating complex swirls with solenoidal mixing (middle cup, medium entropy) until the cup is at equilibrium (right cup, high entropy) and there is just 'brown' coffee – maximally mixed black and white particles. (Aaronson et al., 2014; Carroll, 2017)

Aaronson et al. (2014) model apparent complexity as the Kolmogorov complexity of a coarse-grained representation of the coffee (voxels of coffee color). Initially, the coarse-grained description is simple (white voxels above black voxels). In the intermediate regime, describing the intricate swirls requires many bits, yielding high apparent complexity. Finally, when the coffee is fully mixed, the coarse-grained representation becomes simple again ("all brown voxels"), so apparent complexity is low. Importantly, at the fine-grained voxel level the fully mixed state is maximally complex, but as soon as the system is coarse-grained even slightly, its complexity collapses. This demonstrates that high entropy does not guarantee high complexity and, crucially, the dissociation depends upon the degree of abstraction or coarse graining: i.e., the depth of representation in a hierarchical generative model. The supposition behind the above arguments is that phenomenal richness depends upon the precision (i.e., negative entropy) afforded deep or coarse-grained representations, relative to fine-grained representations at lower hierarchical levels.

If phenomenal content is associated with complexity, then in the early mixing phase, entropy and phenomenological richness rise together. However, beyond a certain point this relationship breaks down: entropy continues to increase, while complexity (and therefore experiential richness) decreases.

A closely related illustrative example is provided by the Ising model across subcritical, critical, and supercritical temperature regimes (McCoy & Wu, 2014). At low temperatures, the system is ordered, with low entropy and low apparent complexity. At high temperatures, the system exhibits high entropy due to microscopic randomness, yet coarse-graining yields a simple, homogeneous description and therefore low apparent complexity. Near the critical temperature, entropy is intermediate, but spatial and temporal long-range correlations emerge across scales, producing maximal apparent complexity under coarse-graining.

This scale-dependence generalizes naturally to generative models of perception. The level of coarse-graining in a generative model is directly related to its number of effective parameters: increasing the number of parameters corresponds to more fine-grained modelling (less coarse-graining), while reducing them enforces a more abstract, coarser description of experience (Liu et al., 2008). In algorithmic information-theoretic terms, one can think of the hierarchical brain as executing a program on a Universal Turing Machine that generates an internal model of the world at successive scales or resolution. On this view, each successive level or scale is trying to compress the scale below and thereby minimize complexity without compromising accuracy. Indeed, one could argue that the notion of deep models inherits from these Bayesian or algorithmic imperatives. Please see (Friston et al., 2021; Hu et al., 2022; Lin et al., 2017; Mehta & Schwab, 2014) for a discussion in the context of generative models and machine learning. In short, complexity isolates the structural aspects of a system by filtering out randomness, i.e., it captures the organized form rather than noise (Gell-Mann & Lloyd, 1996). Following our previous reasoning, higher apparent complexity is related to richer and more elaborate experiences.

Absorptive states such as jhāna or those induced under 5-MeO-DMT increase the level of coarse-graining through suspending attention to the sensorium and thereby biasing inference towards higher levels or scales of representation (see Figure 4). Heuristically, given the coffee cup example, absorptive states would be associated with a coarser sort of visual perception that effectively smooths over fine-grained detail; this is analogous to model underfitting. In these states, the generative model compresses the sensorium (the brain executes a short program in AIT terms), so the complexity is low (because the sensorium is attenuated and therefore there is nothing to compress and the posterior is close to a high entropy prior). HCPEs, by contrast, can be understood as relaxing the coarse-graining imperatives by compressing unattenuated sensory information at lower (fine-grained) scales; thereby increasing complexity. This entails fitting more of the sensory data, manifesting as finer-scale representation, analogous to model overfitting. In AIT terms, the brain is executing a longer program: the complexity is high because the model encodes fine-scale detail. Crucially, the entropy of neuronal representations is always higher than normal—due to imprecise top-down prior constraints—regardless of whether the system is over or underfitting.

This perspective clarifies why—despite a high entropy in both HCPEs and absorptive states—the richness of their phenomenology can be radically different. The difference lies in how coarse-graining shapes complexity and compression: with fine-grained inference (HCPEs overfitting), complexity rises with entropy, until the system becomes saturated with proliferating content; with overly coarse-grained inference (absorptive underfitting), complexity declines because there are no sensory or likelihood constraints. As illustrated in Figure 6, an increase in entropy therefore carries different implications depending on the inferential regime—manifesting as overflow in one case and as minimal-content awareness in the other.

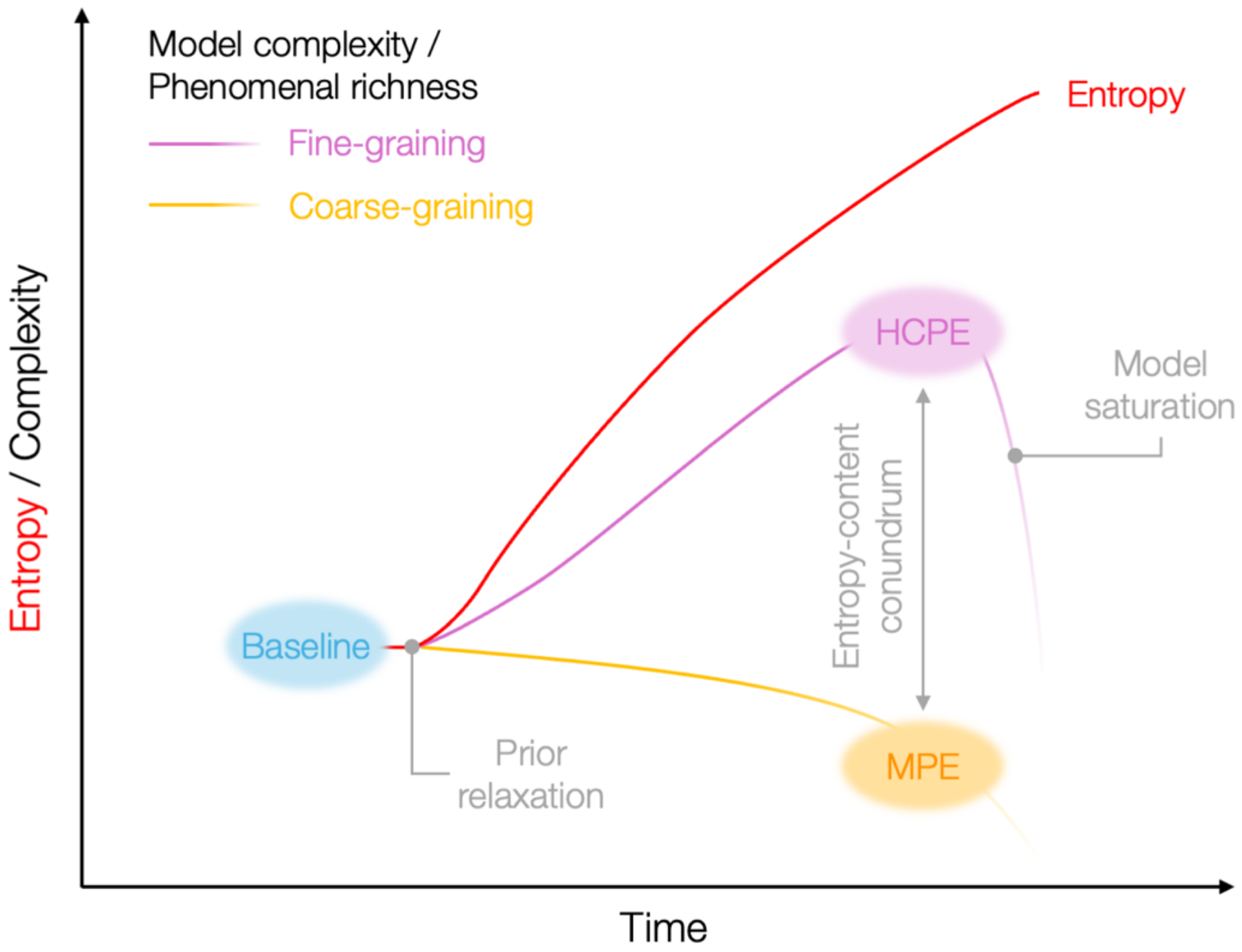

**Figure 6. Entropy rises in both HCPEs and absorptive states, but complexity diverges depending on the inferential regime**. The red curve depicts the entropy of posterior beliefs or their neuronal parameterization; it increases monotonically as prior constraints are relaxed. The purple curve represents complexity under fine-grained HCPE inference: unattenuated sensory fluctuations are amplified into proliferating content, yielding an initial rise and prolonged plateau of high complexity before a late collapse as the model saturates. The yellow curve represents complexity under coarse-grained inference characteristic of Minimal Phenomenal Experience (MPE) states, where experiential content is progressively compressed and complexity decreases monotonically. As time progresses, both HCPEs and absorptive trajectories may converge toward phenomenological cessation at low complexity. This illustrates the divergence between entropy and experiential content across different inferential regimes.

**Expanding the Entropic Brain Hypothesis**

There appears to be a dissociation between states that combine low phenomenal content with high entropy (such as the jhānas) and states that combine low phenomenal content with low entropy (such as anesthesia, or coma) (Figure 7). The increase of entropy with phenomenal richness that is observed from anesthesia to ordinary wakefulness and then to psychedelic states is consistent with the original proposal of the EBH. This suggests that phenomenal richness and entropy are not sufficient to characterize the space of conscious states. Ergo, a third dimension is likely required.

One natural candidate is the level of wakefulness (Figure 7a) (Laureys, 2005), although other interpretations that reflect dynamical properties of the brain are possible, such as neural responsiveness, susceptibility, or proximity to a critical regime. One operational measure that captures this type of property is the Perturbational Complexity Index (PCI), which estimates the degree to which the brain reacts in a structured manner to external perturbations (Casali et al., 2013). PCI is typically low in sleep, anesthesia, and coma, and high during ordinary wakefulness (Casali et al., 2013; Sarasso et al., 2015). Also, PCI remains similarly high in a psychedelic state in comparison to ordinary wakefulness (Ort et al., 2023; Sarasso et al., 2015), although MPEs such as the jhānas still require direct empirical confirmation.

When this three-dimensional structure is projected onto the two-dimensional plane of phenomenal richness and entropy, the resulting visualization reproduces the empirical pattern shown in Figure 7b: MPE and HCPEs have high entropy despite contrasting phenomenology, while sleep appears in the low-richness / low-entropy region.

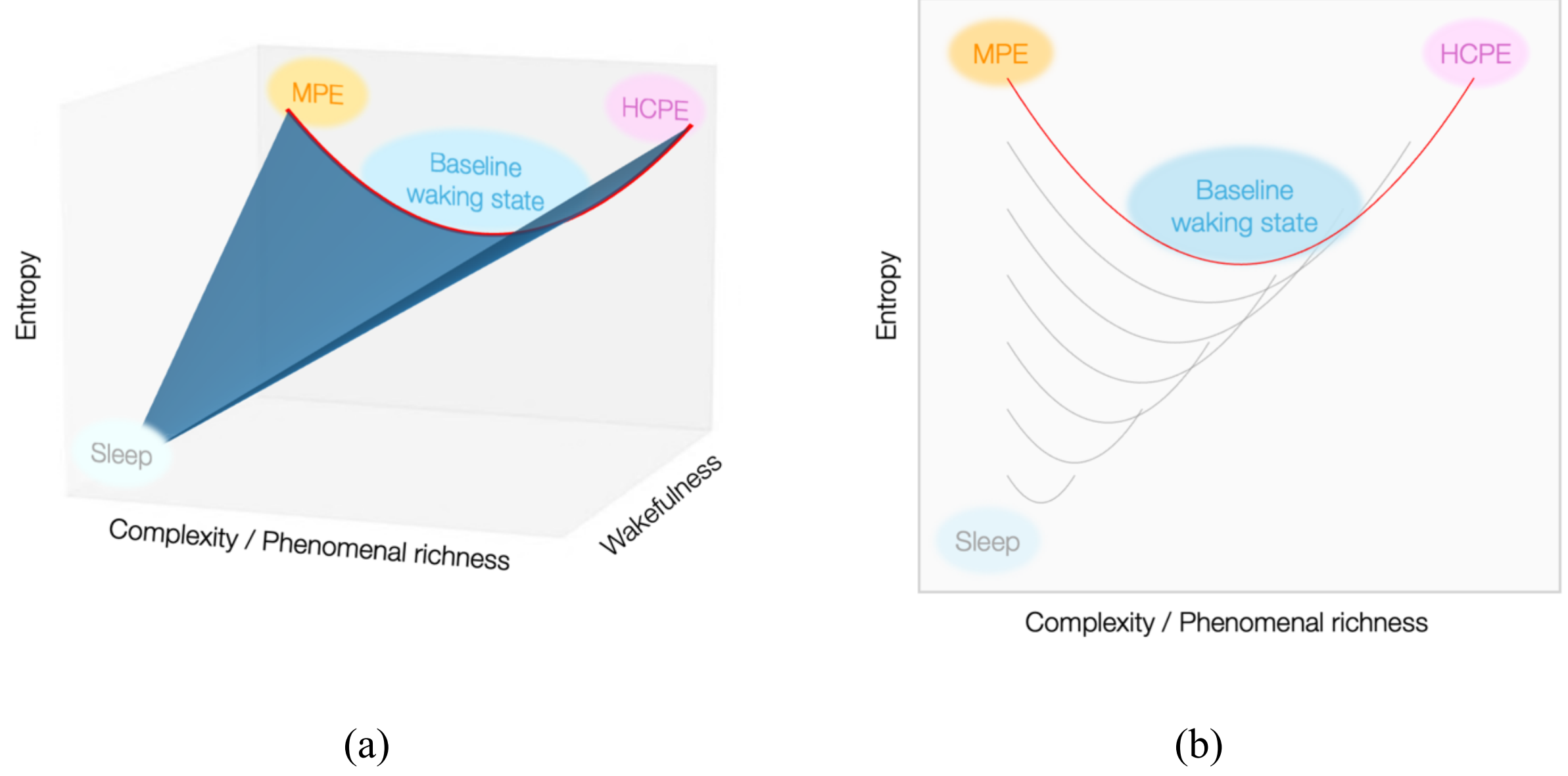

(a) (b)

**Figure 7. Three-dimensional extension of the entropic brain hypothesis. (a)** A three-dimensional schematic showing how adding a third variable, here labeled wakefulness, separates states with low phenomenal richness and high entropy (MPE) or low entropy (sleep). When this surface is projected onto the two-dimensional plane, the pattern in panel (b) emerges. **(b)**

Phenomenal richness vs. entropy. Sleep and anesthesia lie in the region of low richness and low entropy, while HCPEs and MPEs occupy regions of elevated entropy despite opposite phenomenological profiles. The original EBH is marked in blue.

On the present view, this third dimension likely corresponds to the precision afforded to sensory evidence; i.e., the likelihood precision. This is profoundly attenuated in sleep and anesthesia, when there is an effective disengagement from the sensorium (Hobson, 2009; Hobson & Friston, 2012). However, these low levels of consciousness are clearly distinct from absorption states that engender MPE, suggesting that a cessation of hierarchical belief updating (i.e., compression) in anesthesia and coma is distinct from belief updating in absorption states (and possibly dreaming) that rest upon a delicate balance between prior and sensory precision.

**Discussion**

The entropy–content conundrum recognizes that neural entropy is not a simple index of "how rich" someone's experience is. Empirically, both HCPEs and Minimal Phenomenal Experiences (MPEs) occupy regions of liberated inference, despite sitting at opposite extremes of phenomenological richness. The original entropic brain hypothesis captures one aspect of a spectrum where richer experiences tend to co-occur with higher entropy—but it does not by itself explain why phenomenologically simple, absorptive states like jhāna or those induced by 5-MeO-DMT can also exhibit high entropy.

Here, we proposed the Complex Brain Hypothesis (CBH) as a refinement of the EBH. Formally, the CBH distinguishes between the complexity of inference and its accuracy. On this view, phenomenological richness tracks model complexity—how many effective parameters are recruited in the system's "reality model", and with how much weight—whereas entropy reflects the range of plausible explanations that are entertained to explain the sensorium. High entropy can therefore arise in very different inferential regimes: fine-grained, overfitting models that engage too readily with—and potentially amplify—bottom-up sensory data to yield proliferating experiential content (HCPEs), versus coarse-grained, underfitting models that explain attenuated sensory data to yield content-minimal awareness (MPEs).

At a mechanistic level, this distinction can be expressed in terms of precision-weighting and accuracy constraints within hierarchical inference. Both HCPEs and MPEs can be understood as regimes of liberated inference initiated by a reduction in the precision of high-level priors, which increases the entropy (dispersion) of beliefs and their neuronal parameterization. The critical difference is whether the system remains accuracy-constrained by precise sensory evidence. In HCPEs, likelihood precision is relatively preserved (i.e., prediction errors remain influential), so inference must continue to explain the ongoing sensory stream: posterior beliefs are driven away from priors, complexity increases (greater prior–posterior divergence), and residual error is kept low. This is phenomenologically expressed as proliferating, fine-grained content. In MPEs, by contrast, sensory evidence is functionally attenuated (e.g., via sustained absorption/attentional policies), relaxing the accuracy constraint: prediction errors persist but are assigned low precision, posterior beliefs need not track sensory fluctuations, complexity stays low, and higher residual error can be tolerated. This regime is phenomenologically expressed as stillness or content-minimal awareness. This makes the entropy–content dissociation intelligible: entropy can be elevated in both regimes, while phenomenal richness tracks the complexity/accuracy balance rather than entropy alone.

This reconceptualization integrates MPEs into the same computational landscape as HCPEs without forcing them into a single entropy–content axis. It also yields clear empirical and theoretical payoffs. First, it explains how high-entropy neural dynamics can coexist with either experiential overflow or emptiness, depending on the granularity of inference and the position along the underfitting–overfitting continuum. Second, it clarifies the role of hierarchical organization: HCPEs may reflect uncompressed, fine-grained processing that privileges low and mid-level representations, whereas jhāna and similar absorptive states may reflect coarse-grained high or deep representations that are sequestered from the sensorium.

In sum, resolving the entropy–content conundrum requires embedding the entropic brain hypothesis within a more expressive framework that distinguishes entropy from the complexity of inference. The Complex Brain Hypothesis offers such a framework. It preserves the original insight that entropy reflects the dynamical properties of conscious states (e.g., belief updating), while explaining why high entropy can be a signature of both HCPE overflow and minimally structured experiential states. This positions MPEs as decisive test cases for any computational theory that aims to link large-scale brain dynamics with the structure of conscious experience.

To test the CBH, we need to leverage complexity measures in a way that track complexity more closely. While Lempel–Ziv complexity is widely used as a proxy, alternative approaches such as the Block Decomposition Method (Zenil et al., 2018), which has been applied in prior work on the effects of LSD (Ruffini et al., 2023), may offer a better approximation.

**Implications**

The first implication is conceptual. Entropy, as a measure of phenomenological richness, is limited. High entropy can accompany both overflow and emptiness, because both fine-grained inference (HCPEs that amplify minor fluctuations and proliferate content) and coarse-grained inference (absorptive states that do not engage with variability, thus stabilizing awareness) are equivalent in some way, both corresponding to the relaxation of prior precision in the brain. Another interesting difference between MPE and HCPE may be that in MPE there are few belief updates, so that the data is underfitted, but in HCPE there are many updates, thereby overfitting sensory data. Recognizing that entropy can emerge from both fine- and coarse-grained inferential regimes refines and deepens the interpretation of the EBH. The CBH helps accommodate apparent counterexamples (i.e., MPEs) without necessarily doing away with some of the basic principles of the EBH.

A second implication relates to the brain's hierarchical organization in these high-entropy states. In the brain's structural and functional architecture, hierarchical processing unfolds along the unimodal–transmodal gradient (Margulies et al., 2016): unimodal (sensory, posterior) regions implement fine-grained, high-dimensional codes tightly coupled to incoming data; transmodal (association, prefrontal-parietal, default-mode) regions employ coarser, abstract representations integrating information across multiple modalities and timescales (Kiebel et al., 2008). The baseline waking state sits near the optimal point of the hierarchy where local detail and global integration are balanced. Jhāna / 5-MeO states may reflect a collapse of hierarchical differentiation in which representations becomes dominated by transmodal integrative hubs (DMN, precuneus, frontopolar cortex) that attenuate lower-level input, leading to coarse, low-dimensional dynamics but high entropy through global flattening (Carhart-Harris & Friston, 2019). Conversely, HCPEs amplify bottom-up propagation and weaken top-down constraints, driving the system toward a more anarchic quality of organization. High-dimensional, locally decorrelated dynamics may follow—alongside fine-grained but disintegrated processing. In both regimes, entropy rises but for opposite reasons: MPEs through loss of hierarchical differentiation, HCPEs through increased differentiation.

An interesting corollary concerns other forms of contemplative experience and how they might map into this proposed framework. Nondual awareness offers an especially informative case. Prior work has suggested that nondual awareness may involve the system modeling its own generative model (Laukkonen et al., 2025; Lopez-Sola et al., 2024), a configuration that can occur at any level of representational abstraction. From the information-theoretic perspective developed here, this corresponds to the discovery of a highly parsimonious generative model. Nondual awareness would therefore reflect a model with low complexity, and thus one would expect low entropy. Consistent with this interpretation, a recent study, involving an experienced Mahāmudrā practitioner resting in nondual awareness (the recognition of the 'nature of mind'), has found reduced neural complexity as measured with Lempel–Ziv applied to high-definition EEG , although this finding is based on a single participant and should therefore be interpreted as preliminary (Timmermann et al., 2025).

Furthermore, the entropy–content conundrum suggests a useful refinement to theories that frame psychopathology in terms of rigidity, canalization, or disrupted plasticity (Carhart-Harris & Friston, 2019; Juliani et al., 2024; Ruffini et al., 2024). Rather than treating entropy as an unqualified marker of experiential richness, this perspective suggests that phenomenology might also align with the complexity, or "length", of the generative model, while entropy reflects variability around that underlying structure. This shift in emphasis allows psychopathology to be conceptualized as a disorder of model architecture, whether excessively rigid, narrow, or prematurely compressed. In doing so, it complements the aforementioned existing approaches by showing that meaningful therapeutic change additionally depends on reshaping the underlying generative model in ways that restore an appropriate level of complexity and support flexible inference.

Finally, the framework integrates naturally with contemporary computational phenomenology. Metzinger characterizes MPEs as maximally simple, self-knowing states of wakefulness, and suggests that they arise from a nonconceptual representation of tonic alertness, which may constitute the simplest and most basic dimension of experience (Metzinger, 2020, 2024). The present view is compatible with this account, insofar as MPEs are understood as regimes in which belief updating restricted to deep abstract scales, rather than distributed across multiple levels of representation. As a result, phenomenological simplicity coexists with elevated brain entropy, since incoming information is constrained by a single explanatory model, leaving substantial residual variability unresolved rather than elaborated into experiential content.

## Conclusion

Minimal Phenomenal Experiences pose a useful challenge for the entropic brain hypothesis, which links richer phenomenology to higher neural entropy. Empirically, both HCPEs and phenomenologically sparse, absorptive states exhibit elevated signal diversity relative to ordinary wakefulness, despite occupying opposite ends of the experiential richness spectrum. This entropy–content conundrum suggests that entropy alone is insufficient to explain certain conscious states, namely MPEs.

The Complex Brain Hypothesis is proposed as a nuance to the EBH. It distinguishes between the complexity and entropy of representations (where the entropy is a constitutive part of complexity). On this view, phenomenological richness primarily tracks model complexity—how many effective parameters are deployed in the brain's "reality model". High entropy can therefore arise in at least two distinct inferential regimes: 1) an overfitting, fine-grained regime characteristic of many HCPEs; or 2) an underfitting, where there is a low-precision, coarse-grained regime characteristic of MPEs, in which a stable, low-dimensional model smooths away variability into content-minimal awareness.

By embedding entropy within this broader information-theoretic and hierarchical framework, the CBH preserves the core insight of the EBH, while explaining why high entropy can accompany both rich content (HCPE) and low-content (MPE). It predicts that measures sensitive to model architecture and responsiveness—such as approximations of Kolmogorov Complexity, causal large-scale models and carefully controlled neurophenomenological protocols—will be necessary to disambiguate high-entropy states that differ in complexity and phenomenology. In this sense, MPEs are important test cases: any adequate computational theory of consciousness must be able to account for how similarly high-entropy neural dynamics can support both densely structured HCPEs and minimally structured, “contentless” awareness.

**References**

Aaronson, S., Carroll, S. M., & Ouellette, L. (2014). Quantifying the rise and fall of complexity in closed systems: The coffee automaton. *arXiv Preprint arXiv:1405.6903*. https://core.ac.uk/download/pdf/216209378.pdf

Ashby, W. R. (1956). *An introduction to cybernetics*. https://philpapers.org/archive/ASHAIT

Bennett, C. H. (2003). Notes on Landauer's principle, reversible computation, and Maxwell's Demon. *Studies In History and Philosophy of Science Part B: Studies In History and Philosophy of Modern Physics*, *34*(3), 501–510.

Blackburne, G., McAlpine, R. G., Fabus, M., Liardi, A., Kamboj, S. K., Mediano, P. A., & Skipper, J. I. (2025). Complex slow waves in the human brain under 5-MeO-DMT. *Cell Reports*, *44*(8). https://www.cell.com/cell-reports/fulltext/S2211-1247(25)00811-3

Brouwer, A., & Carhart-Harris, R. L. (2021). Pivotal mental states. *Journal of Psychopharmacology*, *35*(4), 319–352.

Carhart-Harris, R. L. (2018). The entropic brain-revisited. *Neuropharmacology*, *142*, 167–178.

Carhart-Harris, R. L. (2025). *The entropic brain today* (Ubzq3_v1). PsyArXiv. https://doi.org/10.31234/osf.io/ubzq3_v1

Carhart-Harris, R. L., & Friston, K. J. (2019). REBUS and the Anarchic Brain: Toward a Unified Model of the Brain Action of Psychedelics. *Pharmacological Reviews*, *71*(3), 316–344. https://doi.org/10.1124/pr.118.017160

Carhart-Harris, R. L., Leech, R., Hellyer, P. J., Shanahan, M., Feilding, A., Tagliazucchi, E., Chialvo, D. R., & Nutt, D. (2014). The entropic brain: A theory of conscious states informed by neuroimaging research with psychedelic drugs. *Frontiers in Human Neuroscience*, *8*, 20.

Carhart-Harris, R. L., Muthukumaraswamy, S., Roseman, L., Kaelen, M., Droog, W., Murphy, K., Tagliazucchi, E., Schenberg, E. E., Nest, T., & Orban, C. (2016). Neural correlates of the LSD experience revealed by multimodal neuroimaging. *Proceedings of the National Academy of Sciences*, *113*(17), 4853–4858.

Carroll, S. (2017). *The big picture: On the origins of life, meaning, and the universe itself*. Penguin.

Casali, A. G., Gosseries, O., Rosanova, M., Boly, M., Sarasso, S., Casali, K. R., Casarotto, S., Bruno, M.-A., Laureys, S., Tononi, G., & Massimini, M. (2013). A Theoretically Based Index of Consciousness Independent of Sensory Processing and Behavior. *Science Translational Medicine*, *5*(198). https://doi.org/10.1126/scitranslmed.3006294

Da Costa, L., Friston, K., Heins, C., & Pavliotis, G. A. (2021). Bayesian mechanics for stationary processes. *Proceedings of the Royal Society A: Mathematical, Physical and Engineering Sciences*, *477*(2256). https://royalsocietypublishing.org/rspa/article/477/2256/20210518/56606

Evans, D. J. (2003). A non-equilibrium free energy theorem for deterministic systems. *Molecular Physics*, *101*(10), 1551–1554. https://doi.org/10.1080/0026897031000085173

Forman, R. K. (1999). *Mysticism, mind, consciousness*. Suny Press. https://books.google.ca/books?hl=en&lr=&id=FO7MOPvo4y8C&oi=fnd&pg=PP11&dq=Forman,+R.+K.+C.+(1990).+Mysticism,+Mind,+Consciousness.&ots=WdsAiMR11h&sig=DxWz2Kar3dc9n06CeNfuTiuM-4A

Friston, K. (2009). The free-energy principle: A rough guide to the brain? *Trends in Cognitive Sciences*, *13*(7), 293–301.

Friston, K. (2010). The free-energy principle: A unified brain theory? *Nature Reviews Neuroscience*, *11*(2), 127–138.

Friston, K., Da Costa, L., Hafner, D., Hesp, C., & Parr, T. (2021). Sophisticated Inference. *Neural Computation*, *33*(3), 713–763. https://doi.org/10.1162/neco_a_01351

Friston, K., Heins, C., Verbelen, T., Da Costa, L., Salvatori, T., Markovic, D., Tschantz, A., Koudahl, M., Buckley, C., & Parr, T. (2025). From pixels to planning: Scale-free active inference. *Frontiers in Network Physiology*, *5*, 1521963.

Friston, K., Kilner, J., & Harrison, L. (2006). A free energy principle for the brain. *Journal of Physiology-Paris*, *100*(1–3), 70–87.

Gell-Mann, M., & Lloyd, S. (1996). Information measures, effective complexity, and total information. *Complexity*, *2*(1), 44–52. https://doi.org/10.1002/(SICI)1099-0526(199609/10)2:1%3C44::AID-CPLX10%3E3.0.CO;2-X

Grünwald, P. D. (2007). *The minimum description length principle*. MIT press. https://books.google.ca/books?hl=en&lr=&id=mbU6T7oUrBgC&oi=fnd&pg=PR9&dq=Gr%C3%BCnwald,+P.+D.+(2007).+The+minimum+description+length+principle.+MIT+press.+https://doi.org/10.7551/mitpress/4643.001.0001&ots=fgoKlnhBkR&sig=EHmAtYNWqLNNCnWAs5CfSj7WDFw

Hinton, G. E., & Van Camp, D. (1993). Keeping the neural networks simple by minimizing the description length of the weights. *Proceedings of the Sixth Annual Conference on Computational Learning Theory  - COLT ’93*, 5–13. https://doi.org/10.1145/168304.168306

Hinton, G. E., & Zemel, R. (1993). Autoencoders, minimum description length and Helmholtz free energy. *Advances in Neural Information Processing Systems*, *6*.

https://proceedings.neurips.cc/paper/1993/hash/9e3cfc48eccf81a0d57663e129aef3cb-Abstract.html

Hobson, J. A. (2009). *The AIM Model of dreaming, sleeping, and waking consciousness*.

Hobson, J. A., & Friston, K. (2012). Waking and dreaming consciousness: Neurobiological and functional considerations. *Progress in Neurobiology*, *98*(1), 82–98.

Hu, H.-Y., Wu, D., You, Y.-Z., Olshausen, B., & Chen, Y. (2022). RG-Flow: A hierarchical and explainable flow model based on renormalization group and sparse prior. *Machine Learning: Science and Technology*, *3*(3), 035009.

Hutter, M. (2005). *Universal Artificial Intelligence*. Springer Berlin Heidelberg. https://doi.org/10.1007/b138233

James, W. (1937). *The Varieties of Religious Experience: A Study in Human Nature; Being the Gifford Lectures on Natural Religion Delivered at Edinburgh in 1901-1902*. Longmans, Green.

Jarzynski, C. (1997). Nonequilibrium Equality for Free Energy Differences. *Physical Review Letters*, *78*(14), 2690–2693. https://doi.org/10.1103/PhysRevLett.78.2690

Jaynes, E. T. (1957). Information Theory and Statistical Mechanics. *Physical Review*, *106*(4), 620–630. https://doi.org/10.1103/PhysRev.106.620

Jaynes, E. T. (1989). The Minimum Entropy Production Principle (1980). In R. D. Rosenkrantz (Ed.), *E. T. Jaynes: Papers on Probability, Statistics and Statistical Physics* (pp. 401–424). Springer Netherlands. https://doi.org/10.1007/978-94-009-6581-2_14

Juliani, A., Safron, A., & Kanai, R. (2024). Deep CANALs: A deep learning approach to refining the canalization theory of psychopathology. *Neuroscience of Consciousness*, *2024*(1), niae005.

Kiebel, S. J., Daunizeau, J., & Friston, K. (2008). A hierarchy of time-scales and the brain. *PLoS Computational Biology*, *4*(11), e1000209.

Klein, M. J., & Meijer, P. H. E. (1954). Principle of Minimum Entropy Production. *Physical Review*, *96*(2), 250–255. https://doi.org/10.1103/PhysRev.96.250

Landauer, R. (1961). Irreversibility and heat generation in the computing process. *IBM Journal of Research and Development*, *5*(3), 183–191.

Laukkonen, R. E., Friston, K., & Chandaria, S. (2025). A beautiful loop: An active inference theory of consciousness. *Neuroscience & Biobehavioral Reviews*, 106296.

Laukkonen, R. E., Sacchet, M. D., Barendregt, H., Devaney, K. J., Chowdhury, A., & Slagter, H. A. (2023). Cessations of consciousness in meditation: Advancing a scientific understanding of nirodha samāpatti. In *Progress in Brain Research* (Vol. 280, pp. 61–87). Elsevier. https://doi.org/10.1016/bs.pbr.2022.12.007

Laureys, S. (2005). The neural correlate of (un) awareness: Lessons from the vegetative state. *Trends in Cognitive Sciences*, *9*(12), 556–559.

Lin, H. W., Tegmark, M., & Rolnick, D. (2017). Why does deep and cheap learning work so well? *Journal of Statistical Physics*, *168*(6), 1223–1247.

Liu, Y., Liang, M., Zhou, Y., He, Y., Hao, Y., Song, M., Yu, C., Liu, H., Liu, Z., & Jiang, T. (2008). Disrupted small-world networks in schizophrenia. *Brain*, *131*(4), 945–961.

Lopez-Sola, E., Sanchez-Todo, R., Vohryzek, J., Castaldo, F., & Ruffini, G. (2024). *An Algorithmic Agent Model of Pure Awareness and Minimal Experiences*. https://files.osf.io/v1/resources/fgxec_v1/providers/osfstorage/66f4804f48fa43651b3cc1ea?action=download&direct&version=3

Mago, J., Brahinsky, J., Miller, M., Maschke, C., Slagter, H. A., Catherine, S., Laukkonen, R. E., Cahn, B. R., Sacchet, M. D., Dixey, W., Dixey, R., Rej, S., & Lifshitz, M. (2025). *Meditative absorption shifts brain dynamics toward criticality* (arXiv:2511.20990). arXiv. https://doi.org/10.48550/arXiv.2511.20990

Maisto, D., Donnarumma, F., & Pezzulo, G. (2015). Divide et impera: Subgoaling reduces the complexity of probabilistic inference and problem solving. *Journal of the Royal Society Interface*, *12*(104). https://royalsocietypublishing.org/rsif/article/12/104/20141335/35413

Margulies, D. S., Ghosh, S. S., Goulas, A., Falkiewicz, M., Huntenburg, J. M., Langs, G., Bezgin, G., Eickhoff, S. B., Castellanos, F. X., Petrides, M., Jefferies, E., & Smallwood, J. (2016). Situating the default-mode network along a principal gradient of macroscale cortical organization. *Proceedings of the National Academy of Sciences*, *113*(44), 12574–12579. https://doi.org/10.1073/pnas.1608282113

McCoy, B. M., & Wu, T. T. (2014). *The two-dimensional Ising model*. Courier Corporation. https://books.google.ca/books?hl=en&lr=&id=YO4xAwAAQBAJ&oi=fnd&pg=PP1&dq=McCoy,+B.+M.,+%26+Wu,+T.+T.+(2014).%C2%A0The+two-dimensional+Ising+model.+Courier+Corporation.+&ots=GUJPjpEK1W&sig=ZiUaqal9pG0H9Ge6JhlHtZYH1fU

Mehta, P., & Schwab, D. J. (2014). *An exact mapping between the Variational Renormalization Group and Deep Learning* (arXiv:1410.3831). arXiv. https://doi.org/10.48550/arXiv.1410.3831

Metzinger, T. (2020). Minimal phenomenal experience: Meditation, tonic alertness, and the phenomenology of “pure” consciousness. *Philosophy and the Mind Sciences*, *1*(I), 1–44. https://doi.org/10.33735/phimisci.2020.I.46

Metzinger, T. (2024). *The elephant and the blind: The experience of pure consciousness: philosophy, science, and 500+ experiential reports*. MIT Press.

Ort, A., Smallridge, J. W., Sarasso, S., Casarotto, S., von Rotz, R., Casanova, A., Seifritz, E., Preller, K. H., Tononi, G., & Vollenweider, F. X. (2023). TMS-EEG and resting-state EEG applied to altered states of consciousness: Oscillations, complexity, and phenomenology. *iScience*, *26*(5), 106589. https://doi.org/10.1016/j.isci.2023.106589

Penny, W. D. (2012). Comparing dynamic causal models using AIC, BIC and free energy. *Neuroimage*, *59*(1), 319–330.

Ramstead, M. J. D., Sakthivadivel, D. A. R., Heins, C., Koudahl, M., Millidge, B., Da Costa, L., Klein, B., & Friston, K. (2022). *On Bayesian Mechanics: A Physics of and by Beliefs* (arXiv:2205.11543). arXiv. http://arxiv.org/abs/2205.11543

Ramstead, M. J. D., Seth, A. K., Hesp, C., Sandved-Smith, L., Mago, J., Lifshitz, M., Pagnoni, G., Smith, R., Dumas, G., Lutz, A., Friston, K., & Constant, A. (2022). From Generative Models to Generative Passages: A Computational Approach to (Neuro) Phenomenology. *Review of Philosophy and Psychology*, *13*(4), 829–857. https://doi.org/10.1007/s13164-021-00604-y

Ruffini, G. (2017). An algorithmic information theory of consciousness. *Neuroscience of Consciousness*, *2017*(1), nix019.

Ruffini, G., Damiani, G., Lozano-Soldevilla, D., Deco, N., Rosas, F. E., Kiani, N. A., Ponce-Alvarez, A., Kringelbach, M. L., Carhart-Harris, R. L., & Deco, G. (2023). LSD-induced increase of Ising temperature and algorithmic complexity of brain dynamics. *PLoS Computational Biology*, *19*(2), e1010811.

Ruffini, G., Lopez-Sola, E., Vohryzek, J., & Sanchez-Todo, R. (2024). Neural geometrodynamics, complexity, and plasticity: A psychedelics perspective. *Entropy*, *26*(1). https://www.mdpi.com/1099-4300/26/1/90

Sajid, N., Parr, T., Hope, T. M., Price, C. J., & Friston, K. (2020). Degeneracy and redundancy in active inference. *Cerebral Cortex*, *30*(11), 5750–5766.

Sakthivadivel, D. A. R. (2022). *Towards a Geometry and Analysis for Bayesian Mechanics* (arXiv:2204.11900). arXiv. https://doi.org/10.48550/arXiv.2204.11900

Sakthivadivel, D. A. R. (2023). *Entropy-Maximising Diffusions Satisfy a Parallel Transport Law* (arXiv:2203.08119). arXiv. https://doi.org/10.48550/arXiv.2203.08119

Sandved-Smith, L., Bogotá, J. D., Hohwy, J., Kiverstein, J., & Lutz, A. (2025). Deep computational neurophenomenology: A methodological framework for investigating the how of experience. *Neuroscience of Consciousness*, *2025*(1), niaf016.

Sarasso, S., Boly, M., Napolitani, M., Gosseries, O., Charland-Verville, V., Casarotto, S., Rosanova, M., Casali, A. G., Brichant, J.-F., & Boveroux, P. (2015). Consciousness and complexity during unresponsiveness induced by propofol, xenon, and ketamine. *Current Biology*, *25*(23), 3099–3105.

Schmidhuber, J. (2010). Formal theory of creativity, fun, and intrinsic motivation (1990–2010). *IEEE Transactions on Autonomous Mental Development*, *2*(3), 230–247.

Seifert, U. (2005). Entropy Production along a Stochastic Trajectory and an Integral Fluctuation Theorem. *Physical Review Letters*, *95*(4), 040602. https://doi.org/10.1103/PhysRevLett.95.040602

Sengupta, B., Stemmler, M. B., & Friston, K. (2013). Information and efficiency in the nervous system—A synthesis. *PLoS Computational Biology*, *9*(7), e1003157.

Sengupta, B., Tozzi, A., Cooray, G. K., Douglas, P. K., & Friston, K. (2016). Towards a neuronal gauge theory. *PLoS Biology*, *14*(3), e1002400.

Seth, A. K., & Bayne, T. (2022). Theories of consciousness. *Nature Reviews Neuroscience*, *23*(7), 439–452.

Solomonoff, R. J. (2009). Algorithmic Probability: Theory and Applications. In F. Emmert-Streib & M. Dehmer (Eds.), *Information Theory and Statistical Learning* (pp. 1–23). Springer US. https://doi.org/10.1007/978-0-387-84816-7_1

Sparby, T., & Sacchet, M. D. (2024). Toward a Unified Account of Advanced Concentrative Absorption Meditation: A Systematic Definition and Classification of Jhāna. *Mindfulness*, *15*(6), 1375–1394. https://doi.org/10.1007/s12671-024-02367-w

Timmermann, C., Sanders, J. W., Reydellet, D., Barba, T., Luan, L. X., Angona, Ó. S., Ona, G., Allocca, G., Smith, C. H., & Daily, Z. G. (2025). Exploring 5-MeO-DMT as a pharmacological model for deconstructed consciousness. *Neuroscience of Consciousness*, *2025*(1), niaf007.

Tomé, T. (2006). Entropy production in nonequilibrium systems described by a Fokker-Planck equation. *Brazilian Journal of Physics*, *36*, 1285–1289.

Tononi, G., Sporns, O., & Edelman, G. M. (1999). Measures of degeneracy and redundancy in biological networks. *Proceedings of the National Academy of Sciences*, *96*(6), 3257–3262. https://doi.org/10.1073/pnas.96.6.3257

Van Lutterveld, R., Chowdhury, A., Ingram, D. M., & Sacchet, M. D. (2024). Neurophenomenological Investigation of Mindfulness Meditation "Cessation" Experiences Using EEG Network Analysis in an Intensively Sampled Adept Meditator. *Brain Topography*, *37*(5), 849–858. https://doi.org/10.1007/s10548-024-01052-4

Varela, F. J. (1996). Neurophenomenology: A methodological remedy for the hard problem. *Journal of Consciousness Studies*, *3*(4), 330–349.

Wallace, C. S., & Dowe, D. L. (1999). Minimum message length and Kolmogorov complexity. *The Computer Journal*, *42*(4), 270–283.

Watson, J. D., Onorati, E., & Cubitt, T. S. (2022). Uncomputably complex renormalisation group flows. *Nature Communications*, *13*(1), 7618.

Winn, J., Bishop, C. M., & Jaakkola, T. (2005). Variational message passing. *Journal of Machine Learning Research*, *6*(4). https://www.jmlr.org/papers/volume6/winn05a/winn05a.pdf?q=variational

Zenil, H., Hernández-Orozco, S., Kiani, N. A., Soler-Toscano, F., Rueda-Toicen, A., & Tegnér, J. (2018). A decomposition method for global evaluation of Shannon entropy and local estimations of algorithmic complexity. *Entropy*, *20*(8), 605.